\documentclass[reprint,prl,twocolumn]{revtex4-1}
\usepackage{mathptmx}
\usepackage{graphicx}
\usepackage{color}
\usepackage{amsmath}
\usepackage{amsfonts}

\renewcommand{\tensor}[1]{\ensuremath{\mathbf{#1}}}
\newcommand{\ket}[1]{\ensuremath{|{#1}\rangle}}
\newcommand{\bra}[1]{\ensuremath{\langle{#1}|}}
\newcommand{\braket}[2]{\ensuremath{\langle {#1} | {#2}\rangle}}
\newcommand{\op}[1]{\ensuremath{\hat{#1}}}
\newcommand{\vect}[1]{\ensuremath{\mathbf{#1}}}
\newcommand{\fig}[1]{Fig.~\ref{#1}}

\graphicspath{{./}{Figures/}}

\begin{document}

\title{Extended Wigner function formalism for the spatial propagation of particles with internal degrees of freedom}
\author{Marcel Utz}
\author{Malcolm H. Levitt}
\affiliation{School of Chemistry, University of Southampton, United Kingdom SO17 1BJ}
\author{Nathan Cooper}
\author{Hendrik Ulbricht}
\affiliation{School of Physics and Astronomy, University of Southampton, United Kingdom SO17 1BJ}
\begin{abstract}
An extended Wigner function formalism is introduced for describing the quantum dynamics of particles with internal degrees of freedom in the presence of spatially inhomogeneous fields. The approach is used for quantitative simulations of molecular beam experiments involving space-spin entanglement, such as the Stern-Gerlach and the Rabi experiment. The formalism allows a graphical visualization of entanglement and decoherence processes. 
\end{abstract}
\date{\today}
\maketitle
The Wigner function formalism~\cite{Wigner:1932tp,Hillery:1984wq,Case:2008vf} provides a compact description of spatial quantum states in terms of a quasi-distribution function in phase space. It  incorporates essential features of the spatial quantum state such as its coherence length and the momentum distribution in a natural manner, and provides an intuitive picture of how the position and momentum distributions evolve in time. In contrast to the classical phase space density, the Wigner quasi-distribution function can exhibit negative values, which are used as a measure for the quantum nature of the state under investigation \cite{leonhardt1997a}.
Quantum states of light \cite{smithey1993measurement} and matter-waves \cite{deleglise2008,kurtsiefer1997measurement} have been characterised by phase-space tomographic homodyne detection, which amounts to  interferometric reconstruction of the Wigner function \cite{bertrand1987tomographic,radon1917berichte}.
The Wigner function is used to calculate interference patterns in matter-wave interferometry experiments~\cite{hornberger2009} and to study decoherence processes \cite{nimmrichter2008theory} as well as quantum carpets \cite{schleich2000a}, with many advantages compared to non-phase-space techniques.

\begin{figure}[b]
\begin{centering}
\includegraphics[width=6cm]{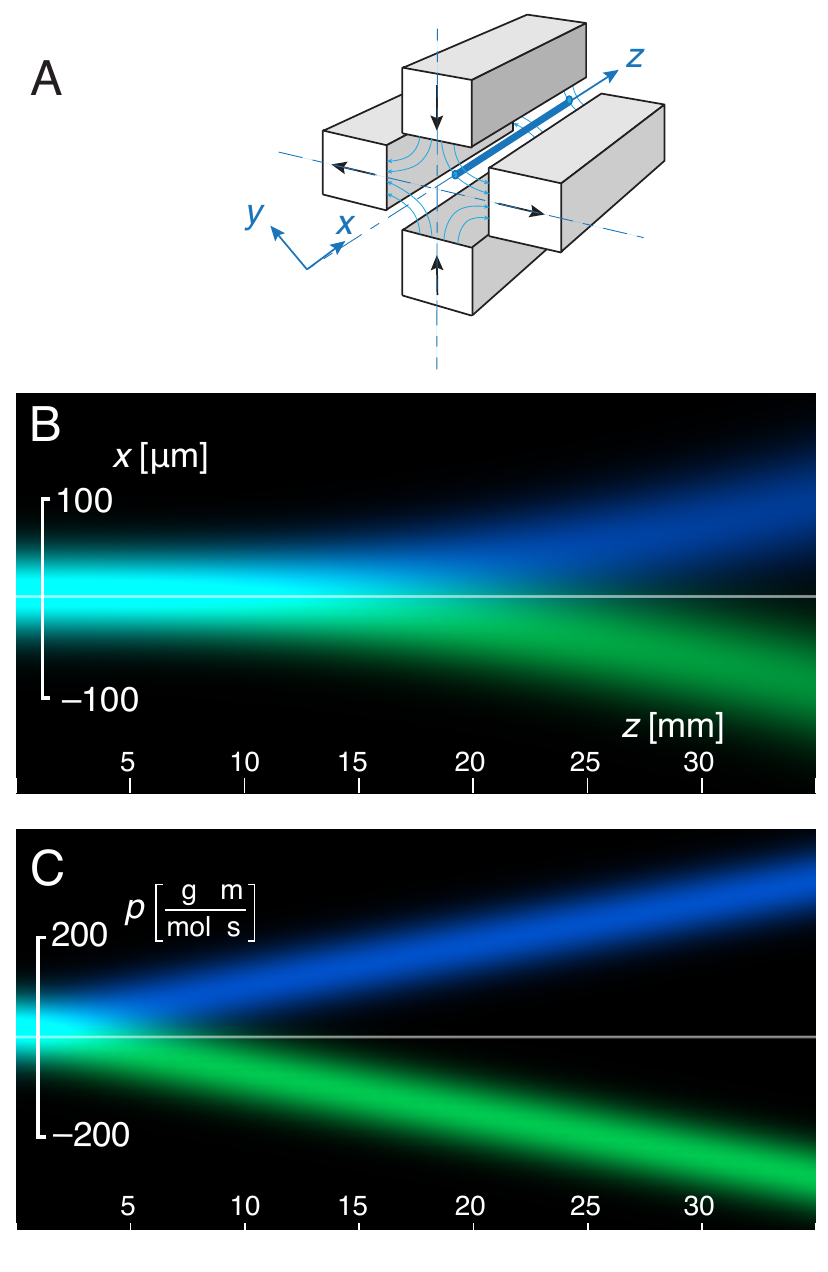}
\caption{A: Magnetic field gradient generated by a quadrupole arrangement of permanent magnets. The beam path indicated (solid blue line) passes through a region where the magnetic field is parallel to the $y$-axis, but varies in magnitude along the $x$-axis (uniaxial field gradient). 
B: Spatial beam trace in the Stern-Gerlach experiment. C: Beam trace in the transverse momentum dimension.}
\label{fig-stern-gerlach-traces}
\end{centering}
\end{figure}

In its current form, the Wigner function formalism is designed for situations in which there is no coupling of the internal degrees of freedom to the spatial propagation. Several generalisations of the Wigner function to other dynamical variables, such as spin, rotation and orientation, have been reported \cite{Chumakov:1999vm,*Klimov:2002wp,*Mukunda:2004tk,*Mukunda:2005vl, *fischer2013}. These approaches treat rotational degrees of freedom through a joint quasi-probability distribution function in angular position and angular momentum space, in direct analogy to the Wigner function treatment of linear position and momentum, but are not capable of describing the coupling of the internal states to the translational motion induced by inhomogeneous external fields. 

The Stern-Gerlach (SG) experiment~\cite{gerlach1922} is a seminal example of a quantum experiment involving coupling between internal and external degrees of freedom. In this experiment, an electron or nuclear spin interacts with a spatially inhomogeneous magnetic field through the magnetic Zeeman interaction. The outcome of the Stern-Gerlach experiment is, of course, ``well-known'': an incident molecular beam of particles with spin-1/2 is separated by the inhomogeneous magnetic field into two beams, each corresponding to particles with well-defined spin angular momenta along the field direction. But how does this separation happen {\em in detail}, on the level of the spatial quantum state?

In this article we present an {\em extended Wigner function} (EWF) which includes the presence of internal degrees of freedom in the propagating particle, and the coupling of those internal degrees of freedom to inhomogeneous external fields.   

An improved modelling of the spatial quantum superposition of particles possessing internal degrees of freedom is highly relevant for predicting the outcome of matter wave diffraction experiments involving particles with vibrational, rotational, and spin degrees of freedom in the presence of inhomogeneous fields~\cite{gring2010,gerlich2008,ulbricht2008,deachapunya2007}. 

The Wigner function \cite{Wigner:1932tp,Hillery:1984wq,Case:2008vf} $W(x,p)$ is a joint quasi-probability density function defined over the combined domains of the spatial coordinate(s) $x$ and its associated momentum (momenta) $p$. It is defined as a Weyl integral transform of the density operator \cite{Fano:1957wf} $\hat{\rho}=\ket{\psi}\bra{\psi}$, of the following form:
\begin{equation}
W(x,p) = \frac{1}{h}\int e^{-\frac{ips}{\hbar}}\,\bra{x+\tfrac{s}{2}}\op{\rho}\ket{x-\tfrac{s}{2}}\;ds.
\end{equation}

Consider a particle with a finite number of internal quantum states. In the discussion below, we refer to these internal states as ``spin states", although the same formalism applies to non-spin degrees of freedom, such as quantized rotational and vibrational states. See the supplementary information for our definition of an internal state~\cite{supp}. We extend the Wigner function by combining it with the density operator formalism commonly used in the quantum description of magnetic resonance~\cite{Fano:1957wf}. The definition of the Wigner function is extended by projecting the density operator onto the spin-state specific position state \ket{x,\eta}, where $\eta=\alpha,\beta,\dots$ denotes the spin state. This results in a Wigner probability density matrix $W_{\eta\xi}(x,p)$, whose elements depend parametrically on the positional variables and their associated momenta:
\begin{equation}
W_{\eta\xi}(x,p) = \frac{1}{h}\int e^{-\frac{ips}{\hbar}}\,\bra{x+\tfrac{s}{2},\eta}\op{\rho}\ket{x-\tfrac{s}{2},\xi}\;ds.
\end{equation}

\begin{figure*}
\begin{centering}
\includegraphics[width=0.8\textwidth]{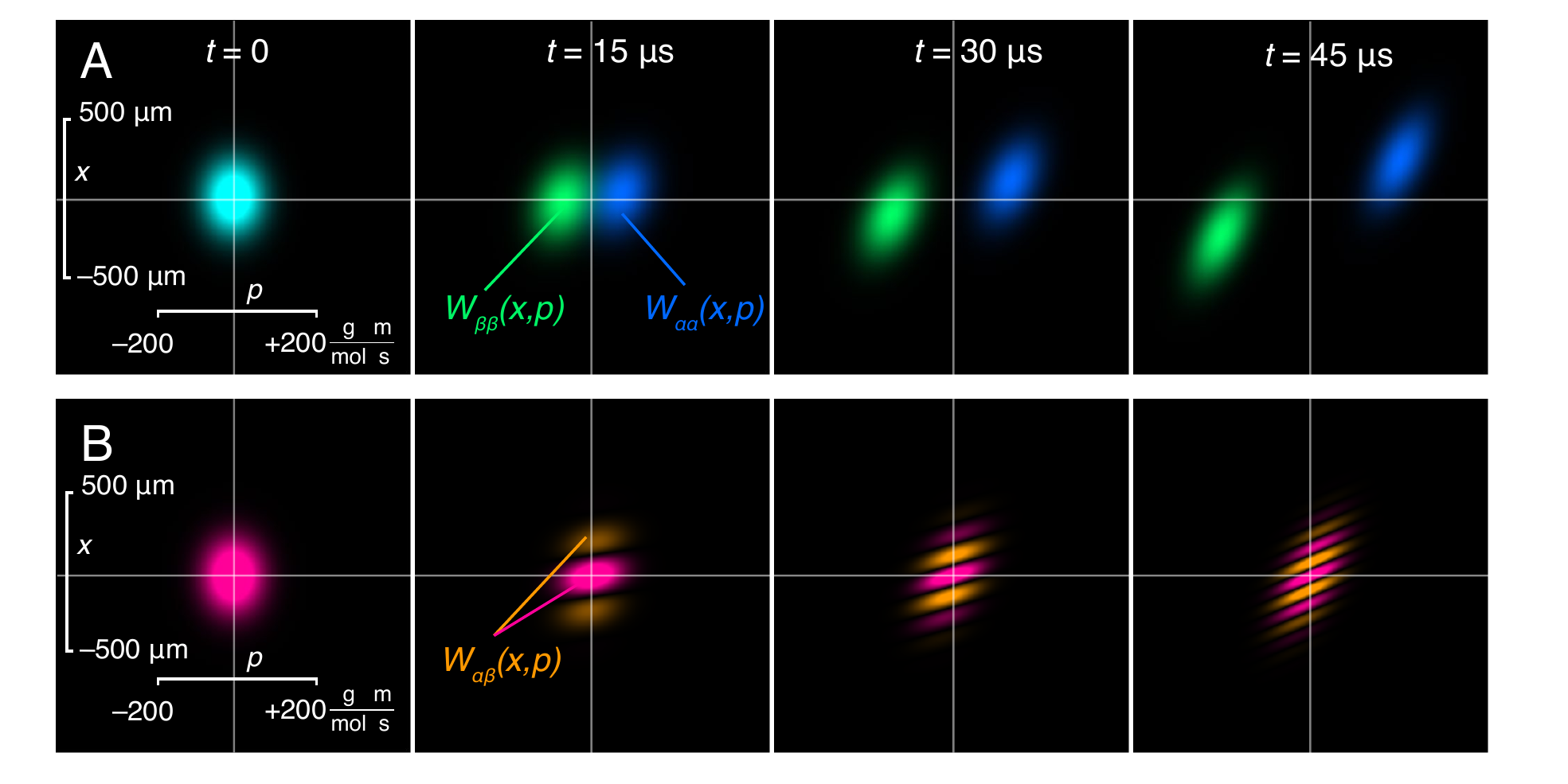}
\caption{A: Evolution of $W_{\alpha\alpha}$ and $W_{\beta\beta}$ under the influence of a magnetic field gradient in a Stern-Gerlach experiment on $Ag$ atoms in a field gradient of 10~$\mathrm{G\,\mu m^{-1}}$, moving at a velocity of 550~$\mathrm{m/s}$ (rms velocity at an oven temperature of $1300$ K). 
B: Evolution of the real part of the off-diagonal element $W_{\alpha\beta}$, assuming a coherent state initially polarised along the $x$-axis. The strength of the magnetic field gradient has been reduced by a factor of $5\times10^4$ compared to A in order to make the spatial modulation visible. The shearing of the fine structure of the Wigner function represents decoherence.
}
\label{fig-sg-Waabb}
\end{centering}
\end{figure*}

An extended Wigner function of this type was defined by Arnold and Steinr\"uck~\cite{arnold1989} , but without elucidating its application to the simulation of quantum dynamics. Our main interest lies on the quantum dynamics of particles in the presence of spatially inhomogeneous and possible time-dependent fields which couple to the spin degrees of freedom. For non-relativistic and uncharged particles, the Hamiltonian may be written as
\begin{equation}
\op{H} = \frac{\op{p}^2}{2m}+ U(x,\op{S}),
\end{equation}
where $x$ is the position, and $\op{p}$ and $\op{S}$ denote the operators associated with the momentum, and spin degrees of freedom, respectively.
It is convenient to consider the contributions of the kinetic and potential energy parts of the Hamiltonian to the time derivative of the Wigner functions separately.
 It can be shown through integration by parts that the kinetic energy contribution is
\begin{equation}\label{eq-WdotT}
\left[\dot W_{\eta\xi}(x,p)\right]_T = -\frac{p}{m} \frac{\partial}{\partial x} W_{\eta\xi}(x,p),
\end{equation}
while the potential energy part of the Hamiltonian contributes as follows:
\begin{equation}\label{eq-pot-evolution-commuting}
\begin{split}
\left[\dot W_{\eta\xi}(x,p)\right]_U = \\ 
\frac{1}{i\hbar}\sum\limits_n\frac{1}{n!}\left(\frac{\hbar}{2i}\right)^n\frac{\partial^n W_{\eta\xi}}{\partial p^n}\left[ (-)^n  \frac{\partial^n U_{\eta\eta}}{\partial x^n} 
- \frac{\partial^n U_{\xi\xi}}{\partial x^n}
\right].
\end{split}
\end{equation}
In this expression, the basis of the spin degrees of freedom has been chosen to diagonalise the potential energy: $U_{\eta\xi}(x)=\delta_{\eta\xi} \bra{x,\eta}U(x,\op{S})\ket{x,\xi}$. Obviously, this is only possible if the potential energy operator at different locations commutes: $[U(x_1,\op{S}),U(x_2,\op{S})]=0\;\forall x_1,x_2$. An expression corresponding to \eqref{eq-pot-evolution-commuting} for the general case is given in the supplementary material.

The series \eqref{eq-pot-evolution-commuting} converges rapidly if the coherence length $l_c$ of the quantum state represented by the Wigner function is short compared to the length scale of variation of $U(x,\op{S})$. In the momentum dimension, the Wigner function typically has a Gaussian shape of width $\hbar/l_c$, and the derivatives ${\partial^n W_{\eta\eta}}/{\partial p^n}$ scale with $(l_c)^{n}$. By contrast, the spatial derivatives of a harmonic potential with period $L$ scale with $L^{-n}$. Together, the terms in \eqref{eq-pot-evolution-commuting} therefore scale as $(l_c/L)^n/n!$. 
If $l_c\ll L$, \eqref{eq-pot-evolution-commuting} may be truncated to first order, yielding
\begin{equation} \label{eq-Wdot-o1}
\left[\dot W_{\eta\xi}(x,p)\right]_U =  W_{\eta\xi}\frac{U_{\eta\eta}-U_{\xi\xi}}{i\hbar}-\frac{\partial W_{\eta\xi}}{\partial p}\frac{F_{\eta\eta}+F_{\xi\xi}}{2},
\end{equation}
where $F_{\eta\eta}(x)= - \partial U_{\eta\eta}/\partial x$ is the force acting on the quantum state $\eta$.
Together with the evolution due to kinetic energy, this set of partial differential equations can be integrated numerically, forming the basis of detailed simulations of the quantum state propagation in the presence of inhomogeneous fields. In the form given above, which assumes a diagonal Hamiltonian, the different elements of the Wigner matrix $(W_{\eta\xi})$ are decoupled, and therefore evolve independently from each other. If the Hamiltonians in different positions do not commute, however, the full version of \eqref{eq-pot-evolution-commuting} applies, which couples the internal states. See the supplementary information for a derivation and discussion of eqns. \eqref{eq-WdotT} and \eqref{eq-Wdot-o1}~\cite{supp}.

We now illustrate the application of both equations to some molecular beam experiments.

In the Stern-Gerlach experiment, a beam of spin-$1/2$ particles  is exposed to a lateral magnetic field gradient. We define the axis of the molecular beam apparatus as $z$, and assume that the magnetic field varies in the transverse $x$-direction. The potential energy part of the Hamiltonian in the presence of an external magnetic field $\vect{B}$ is then given by
\begin{equation}
U(\op{\vect{S}},x)= -\hbar\gamma\vect{B}(x)\cdot\op{\vect{S}}.
\end{equation}

The original magnet design used  by Stern and Gerlach~\cite{gerlach1922} produces divergent magnetic field  lines at the location of the beam. This corresponds to a biaxial magnetic field gradient tensor, requiring two spatial dimensions to be included in the Wigner function. To avoid this complication, we use a different arrangement, in which the magnetic field gradient is uniaxial. In this case, the magnetic field lines are all parallel, but vary in density in the direction perpendicular to the magnetic field itself. Magnetic fields of this type occur in quadrupole polarisers, as shown in \fig{fig-stern-gerlach-traces}A.

We assume the magnetic field points along the $y$-axis, and varies linearly in magnitude along the $x$-axis, $
\tensor{B}(x,y,z) = \left( B_{y0} + x\,G_{xy}\right)\,\tensor{e}_y,
$
where $B_{y0}$ is the magnetic field at $x=0$, and $G_{xy}=\partial B_y/\partial x$. This field is fully consistent with Maxwell's equations, since it satisfies $\nabla\cdot \tensor{B}=0$. The field gradient has only a single non-zero cartesian component $\label{eq-uniaxial-field-gradient} \nabla\tensor{B} = G_{xy}\;\tensor{e}_x\tensor{e}_y .$ We choose the spin states \ket{\alpha} and \ket{\beta} as the eigenstates of $\op{S}_y$, such that the matrix elements of the potential part of the Hamiltonian are
\begin{equation}\label{eq-diagonal-potential-hamiltonian}
\begin{array}{ll}
U_{\alpha\alpha}(x) = -\frac{\gamma\hbar}{2}B_y(x) & 
U_{\alpha\beta}(x)  = 0 \\[2mm]
U_{\beta\alpha}(x) = 0  & 
U_{\beta\beta}(x)  = +\frac{\gamma\hbar}{2}B_y(x).
\end{array}
\end{equation}
The resulting equations of motion for the EWF matrix elements are given in the SI.

In its original form, the Stern-Gerlach experiment was conducted on a beam of Ag atoms emanating from an oven at a temperature of about 1300 K. The magnetic field gradient was of the order of 10~G/cm over a length of 3.5~cm \cite{Friedrich:2003td}.  For simplicity, we ignore the nuclear spin of Ag, and treat the atoms as (electron) spin 1/2 particles. In the case of magnetic fields larger than the hyperfine splitting (about 610~G \cite{Wessel1953ha} in the case of Ag), this is a good approximation, since the nuclear and the electron spin states are essentially decoupled. The root mean square velocity of Ag atoms at 1300~K is approximately 550~m/s.  After leaving the oven, the Ag atoms are collimated by a pair of collimation slits $30\;\mathrm{\mu m}$ wide and separated by 3~cm. The longitudinal momentum of the silver atoms is approximately $6\times10^4\;\mathrm{g\,mol^{-1}\,ms^{-1}}$. The collimation aspect ratio of 1:1000 therefore
results in a transverse momentum uncertainty of $\Delta p=60\;\mathrm{g\,mol^{-1}\,ms^{-1}}$, which corresponds to a 30~$\mu$m wide beam with a transverse coherence length of about $l_c=h/\Delta p\approx 7$~nm.

An unpolarised beam entering the magnetic field gradient is represented by a unity spin density matrix, such that
$
W_{\alpha\alpha}(t=0)=W_{\beta\beta}(t=0)=W_0(x,p),
$
where the initial state $W_0(x,p)$ is a two-dimensional normalised Gaussian function centred at $(x,p)=(0,0)$, with widths given by
 coherence length $l_c$ and the beam width $\Delta x$ (cf. SI). The off-diagonal Wigner functions vanish: $W_{\alpha\beta}=W_{\beta\alpha}\equiv 0$, and the diagonal ones can be obtained in closed form by integrating the equations of motion (cf. SI).

\begin{figure}
\begin{center}
\includegraphics[width=7cm]{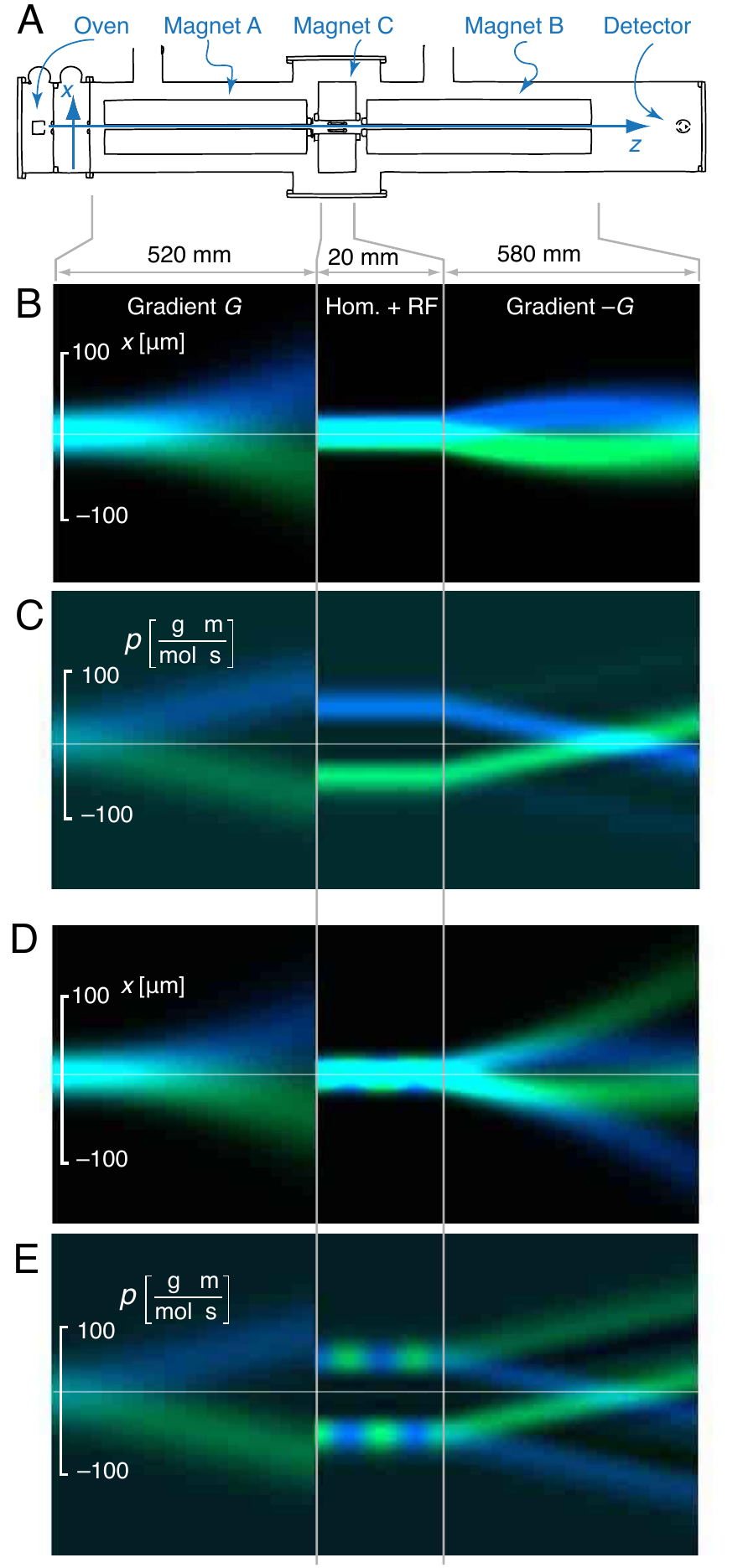}
\caption{\label{fig-rabi-traces}
A: Schematic of the original Rabi apparatus (adapted from \cite{Rabi:1939cb}). B, C:
Simulated spatial (B) and momentum (C) beam traces for $B_{y0}$ far from the resonance condition.  D, E Simulated spatial (D) and momentum (E) traces at the magnetic resonance condition. The intensity scale has been multiplied by a factor of 10 to the right of Magnet A due to the reduction in total beam intensity by the collimation slit between Magnets A and C.
}
\end{center}
\end{figure}

\begin{figure}
\begin{center}
\includegraphics[width=4.5cm]{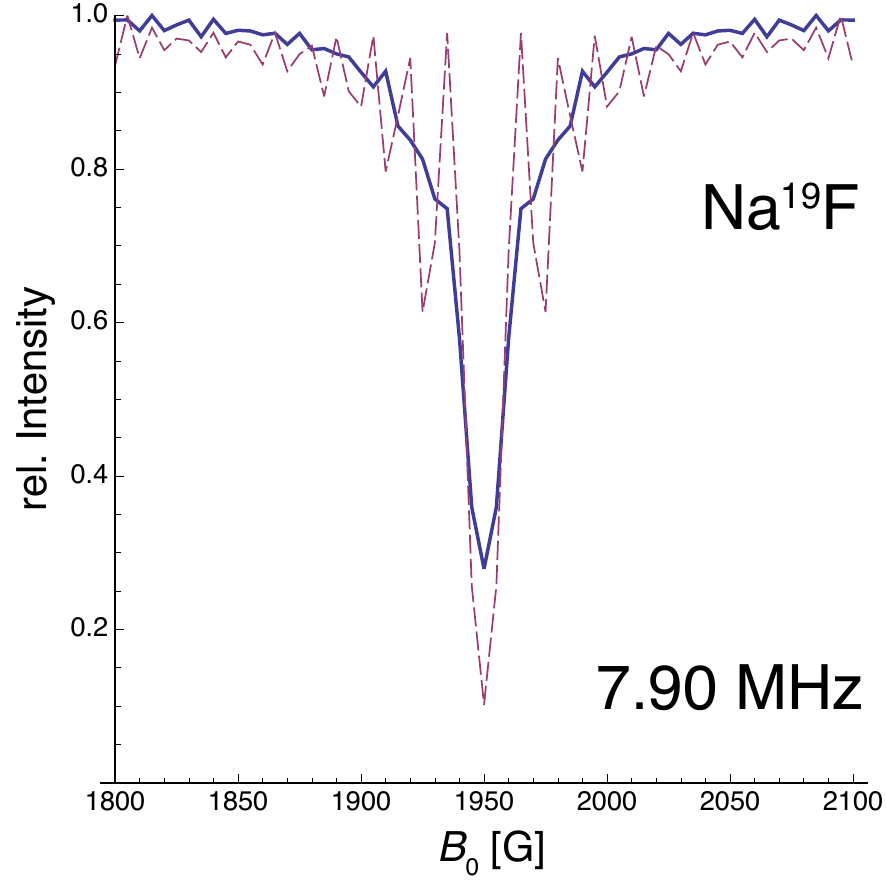}
\caption{Predicted beam intensity in the Rabi experiment as a function of magnetic field $B_0$ in the homogeneous section. Dashed line: single velocity $v=500\;\mathrm{m/s}$; solid line: average over 4 different velocities from 450 to $550\;\mathrm{m/s}$.}
\label{fig-NaF-spectrum}
\end{center}
\end{figure}
 
\fig{fig-stern-gerlach-traces}B shows the projections of the Wigner matrix elements $W_{\alpha\alpha}$ and $W_{\beta\beta}$ onto the spatial axis as a function of position along the beam path in blue and green, respectively. The initially unpolarised beam begins to split after about 10~mm, and is completely separated after 25~mm. As expected, the separation of the two beams grows quadratically along the beam path. The corresponding projection onto the momentum dimension is shown in \fig{fig-stern-gerlach-traces}C. Due to the constant, equal and opposite forces experienced by the two polarisation states, the transverse momentum grows linearly along the beam path. It is interesting to note that in the momentum dimension, the beam is fully polarised beyond 5~mm, while spatial separation does not occur until 25~mm. This is also reflected in the Wigner function ``snapshots'' shown in \fig{fig-sg-Waabb}A. In these panels, the transverse momentum and position are plotted on the horizontal and vertical axes, respectively. The beam is initially unpolarised and centred. Under the influence of the field gradient, it splits into two separate spots in the momentum direction first, which gradually drift apart in the position dimension, as well. The peaks of the two distributions $W_{\alpha\alpha}$ and $W_{\beta\beta}$ describe parabolic trajectories in the $x,p$-plane in opposite directions. The evolution of the Wigner matrix elements also shows the gradual shearing due to ballistic drift, which leads to divergence of the beams. 
It should be noted that the final separation of the beams at $z=3.5$~cm amounts to about 200~$\mu$m, which is in quantitative agreement with Stern and Gerlach's observation.

The original Stern-Gerlach apparatus has inspired a substantial number of related experimental arrangements. In particular, the Stern-Gerlach {\em interferometer} \cite{Wigner:1963tw,Scully:1989wd} is of interest in the present context. It relies on the separation and subsequent interference of a coherent spin state in a pair of magnetic field gradients of opposite polarity. While the complete simulation of such a system is outside the scope of this letter, it is instructive to contrast the fate of a coherent spin state with the evolution of the unpolarised beam discussed above. 

Instead of an unpolarised Ag beam, consider one that has been fully polarised in the $x$ direction before entering the field gradient shown in \fig{fig-stern-gerlach-traces}A. This could be accomplished, for example, by preceding the magnet with a similar one rotated by 90$^\circ$ about the $z$-axis, and selecting one of the two resulting traces. 

Polarisation along the $x$-axis corresponds to a spin quantum state $2^{-1/2}(\ket\alpha+\ket\beta)$, and the initial conditions for the EWF matrix elements are then
$W_{\alpha\alpha}=W_{\beta\beta}=W_{\alpha\beta}=W_{\beta\alpha}=\frac{1}{2}W_0$. While the diagonal EWF matrix elements evolve as discussed above, the {\it off-diagonal elements} behave differently. Under the potential energy term \eqref{eq-diagonal-potential-hamiltonian}, they undergo a harmonic oscillation with a linearly position-dependent frequency. This leads to a spatial modulation with wave number $k=\pm \gamma G_{xy} t$. At the same time, however, the ballistic drift shears the Wigner function. Therefore, the direction of the phase modulation in the $(x,p)$-plane gradually rotates, and the Wigner function is modulated in {\em both} the position and momentum domain. This is shown in \fig{fig-sg-Waabb}B. It should be noted that the spatial frequency of the modulation grows very quickly as a function of time; the field gradient used for the simulation shown in \fig{fig-sg-Waabb}B was lowered by a factor of $5\times 10^4$ compared to \fig{fig-sg-Waabb}A in order to make the modulation visible. Under the true field gradient in the SG experiment (10~$\mathrm{G\;\mu m^{-1}}$), the spatial frequency of modulation after $45\;\mathrm{\mu s}$ would already be 1260~$\mathrm{\mu m^{-1}}$. 
It is important to note that due to the shear of the EWF due to ballistic drift, the projections of this modulated EWF on either the momentum or the position axis vanish. Further simulations show that in principle, the coherence may be retrieved by applying a sequence of gradients in the opposite sense, providing that the coherence length of the particle is sufficiently large in the longitudinal direction.

As a second example, we treat the classic magnetic resonance experiment introduced by Rabi and coworkers in order to measure nuclear gyromagnetic ratios \cite{Rabi:1939cb}. The apparatus is shown in \fig{fig-rabi-traces}A.  It relies on two magnetic field gradients of opposite polarity (Magnets A and B). The first gradient imparts a curvature to the beam path depending on the spin state of the entering particle. This curvature is reversed in the second gradient, thus refocusing the beam. In between the two sets of gradients, there is a region with a  homogeneous static magnetic field $B_{y0}$ (Magnet C), combined with a radio frequency field $B_x(t)=B_1 \cos({\omega_\text{rf}\, t})$. 

The spins undergo nutations at a frequency proportional to $B_1$, if $\omega_\text{rf}$ is sufficiently close to the Larmor frequency $\omega = -\gamma B_{y0}$, where $\gamma$ denotes the gyromagnetic ratio. This nutation interferes with the refocusing of the beam, and leads to a measurable decrease in the detected beam intensity. The gyromagnetic ratio can then be inferred from the frequency $\omega_{\rm rf}$ at which the effect is maximal. 

Using the EWF formalism, it is straightforward to simulate this experiment. We have assumed the beam to consist of NaF molecules emanating from an oven at 1300~K. The molecules are treated as single spin 1/2 systems with a gyromagnetic ratio corresponding to $\mathrm{^{19}F}$; the Na nuclear spin is ignored. The geometry of the apparatus and the magnitudes of the magnetic fields and field gradients have been taken from ref.~\cite{Rabi:1939cb}. The evolution of the EWF matrix elements has been computed numerically by Runge-Kutta integration of the equations of motion. The EWF were represented by a structured finite element mesh using bilinear interpolation. The full EWF matrix elements are given in the supplement~\cite{supp}.
 
\fig{fig-rabi-traces}B and C show the position and momentum traces in the case of a large resonance offset. The first magnet splits the beam in a manner analogous to the Stern-Gerlach experiment. A narrow collimation slit then admits only the centre of the beam to the homogeneous magnet region. As a result, the beam entering C is unpolarised in the spatial domain, but completely polarised in the momentum direction (i.e., the transverse momentum and spin states are entangled). The two beams retain their spin ``identity'', and are then spatially refocused by the inverse field gradient (Magnet B). The situation is different when the magnetic field $B_{y0}$ is close to resonance (\fig{fig-rabi-traces}D and E). 
The spin states are now exchanged periodically under the influence of the resonant radio frequency field in Magnet C. As a result, a large fraction of the beam intensity is further deflected by the refocusing magnet, leading to a decrease of the detector signal.  

\fig{fig-NaF-spectrum} shows the computed beam intensity at the detector as a function of $B_{y0}$, assuming an rf frequency and amplitude of 7.90~MHz and 20~G, respectively. Assuming a single velocity of the NaF molecules leads to a sinc-shaped resonance line (dashed line), which is smoothed out if the results are averaged over a 20\% velocity variation. This results in a line shape that is very similar to the ones reported in the original work by Rabi {\em et al} \cite{Rabi:1939cb}. 

In summary, we have introduced an extension of the Wigner function formalism to particles with internal (spin) degrees of freedom. We have demonstrated how this approach can be used to obtain detailed simulations of the quantum dynamics in experiments that rely on the entanglement of spatial and spin parts of the quantum state. The extended Wigner function allows a compelling graphical visualization of quantum superposition and decoherence processes. Future work will be concerned with using this new tool to model closed spin-matterwave interferometers, such as the Humpty-Dumpty experiment~\cite{Scully:1989wd}, for macroscopic quantum superposition experiments. In a future application the EWF could be used to analyse detailed properties of complex quantum systems, such as large molecules.  For example in~\cite{gring2010} the behaviour of the internal state population is mapped onto the centre of mass motion of the molecules by an inhomogeneous electric field. The EWF approach could also be used to broaden the applicability of phase-space descriptions of the dynamics of ultracold atoms, such as those applied to atoms confined in optical lattices in \cite{scott, morinaga, kanem}, by allowing the atoms' multiple internal states to be included in the model. This would be particularly useful when considering effects that rely on the coupling of the atoms' internal states to their translational motion, such as Sisyphus cooling \cite{dalibard}. Last not least the EWF can be applied to model magnetic resonance techniques capable of detecting a small number of spins. Applications to spatially multiplexed NMR experiments such as ultrafast 2D-NMR~\cite{frydman2002} may also be envisaged. 

{\it Acknowledgments.} -- HU would like to thank for support: EPSRC (EP/J014664/1), the Foundational Questions Institute (FQXi), and the John Templeton foundation (grant
39530). This research was supported by the European Research Council.

\begin{widetext}
\section*{Supplementary information}
Here we derive the general propagation equation including spin, give the explicit form of the Gaussian function we used to describe the particle beam profile, and we give the potential energy matrices for the Stern-Gerlach and Rabi experiments. 

\section{General Propagation Equation}
For a particle with spin, the definition of the Wigner function can be extended by projecting the density operator onto the spin-state specific position state \ket{x,\eta}, where $\eta=\alpha,\beta,\dots$ denotes the spin state. This leads to the Wigner matrix
\begin{equation}
W_{\eta\xi}(x,p) = \frac{1}{h}\int e^{-\frac{ips}{\hbar}}\,\bra{x+\tfrac{s}{2},\eta}\op{\rho}\ket{x-\tfrac{s}{2},\xi}\;ds.
\end{equation}
The time evolution of the Wigner matrix is given by the Schr\"odinger equation
\begin{equation}
\frac{\partial}{\partial t} \ket{\psi} = \frac{1}{i\hbar} \left[ \frac{\op{p}^2}{2m}+ U(\op{x},\op{S})\right] \ket{\psi},
\end{equation}
where the potential energy $U$ depends on position \op{x} and spin \op{S}, and \op{p} is the momentum operator. In order to facilitate actual calculations, we express the Schr\"odinger equation in the spin-localised basis \ket{x,\eta}, with the wave functions $\psi_\eta(x)=\braket{x,\eta}{\psi}$. This yields a system of differential equations
\begin{equation}\label{eq-schroedinger}
\frac{\partial}{\partial t}\psi_\eta(x) = -\frac{\hbar}{2im}\frac{\partial^2} {\partial x^2} \psi_\eta(x) + \frac{1}{i\hbar} U_{\eta\xi}(x) \psi_\xi(x),\vspace{2mm}
\end{equation}
where the potential energy is expressed through the matrix elements
$
U_{\eta\xi}(x) = \bra{x,\eta} \op{U} \ket{x,\xi}.
$
Note that we use the Einstein convention, i.e., 
\begin{equation}
U_{\eta\xi}\psi_\xi = \sum\limits_{\xi=\alpha,\beta,\dots}U_{\eta\xi}\psi_\xi.
\end{equation}
From its definition, the time derivative of the Wigner matrix is
\begin{equation}
\dot W_{\eta\xi}(x,p) = \frac{1}{h}\int e^{-\frac{i p s}{\hbar}} \dot\psi_\eta(+)\psi_\xi^\ast(-)+\psi_\eta(+)\dot\psi_\xi^\ast(-)\, ds,
\end{equation}
where we have used the shorthand $\psi_\eta(\pm)=\psi_\eta(x\pm\tfrac{s}{2})$.
The time derivatives of the wave functions are given by the Schr\"odinger equation \eqref{eq-schroedinger}. It is convenient to deal with the kinetic and potential energy terms separately. For the kinetic term, it can be shown through integration by parts that
\begin{equation}\label{eq-WdotT}
\left[\dot W_{\eta\xi}(x,p)\right]_T = -\frac{p}{m} \frac{\partial}{\partial x} W_{\eta\xi}(x,p).
\end{equation}
This expresses ballistic drift through phase space, equivalent to classical mechanics and is Eqn.4 in the paper. It amounts to the shearing transformation
\begin{equation}
W_{\eta\xi}(x,p;t+\delta t)= W_{\eta\xi}(x-\frac{p}{m}\delta t,p ; t).
\end{equation}
The evolution due to the potential energy term is more cumbersome. We have
\begin{equation}\label{eq-WdotU}
\begin{aligned}
\left[\dot W_{\eta\xi}(x,p)\right]_U = \\
=\frac{2\pi}{ih^2}\int e^{-\frac{i p s}{\hbar}}\left[\psi_\xi^\ast(-)U_{\eta\zeta}(+)\psi_\zeta(+)   \right. \\
\left. -\psi_\eta(+) U_{\xi\zeta}(-)\psi^\ast_\zeta(-) \right] ds,
\end{aligned}
\end{equation}
where we have made use of $U_{\zeta\xi}^\ast = U_{\xi\zeta}$. The matrix elements of the potential energy are expanded into a Taylor series about $x$:
\begin{equation}
U_{\xi\zeta}(x\pm\tfrac{s}{2}) = \sum\limits_{n=0}^\infty \frac{1}{n!}\frac{\partial^n U_ {\xi\zeta}(x)}{\partial x^n} \left(\pm\frac{s}{2}\right)^n.
\end{equation}
From the definition of the Wigner matrix, we note furthermore that
\begin{equation}
\left(\pm\frac{\hbar}{2i}\right)^n \frac{\partial^n W_{\xi\zeta}}{\partial p^n} = \frac{1}{h}\int e^{-\frac{i p s}{\hbar}} \left(\mp\frac{s}{2}\right)^n \psi_\xi(+)\psi^\ast_{\zeta}(-)\;ds.
\end{equation}
Inserting this into \eqref{eq-WdotU}, we obtain

\begin{equation}
\left[\dot W_{\eta\xi}(x,p)\right]_U = \frac{1}{i\hbar}\sum\limits_n\frac{1}{n!}\left(\frac{\hbar}{2i}\right)^n\left[ (-)^n  \frac{\partial^n U_{\eta\zeta}}{\partial x^n} \frac{\partial^n W_{\zeta\xi}}{\partial p^n} 
- \frac{\partial^n W_{\eta\zeta}}{\partial p^n}\frac{\partial^n U^\ast_{\zeta\xi}}{\partial x^n}
\right].
\end{equation}

If the basis for the internal degrees of freedom is chosen such that the spin Hamiltonian is diagonal, we have $U_{\eta\xi} = \delta_{\eta\xi} U_{\eta\eta}$, and the above expression simplifies slightly to
\begin{equation}\label{eq-pot-evolution-commuting}
\left[\dot W_{\eta\xi}(x,p)\right]_U = \frac{1}{i\hbar}\sum\limits_n\frac{1}{n!}\left(\frac{\hbar}{2i}\right)^n\frac{\partial^n W_{\eta\xi}}{\partial p^n}\left[ (-)^n  \frac{\partial^n U_{\eta\eta}}{\partial x^n} 
- \frac{\partial^n U_{\xi\xi}}{\partial x^n}
\right].
\end{equation}
It should be noted, however, that this choice is only possible if the spatial dependence is such that the Hamiltonian at different positions commutes. If this is not the case, techniques such as the superadiabatic formalism of Berry may be used to obtain approximate expressions for the propagator~\cite{berry1987,deschamps2008}.

This series converges rapidly in most practical cases. For example, consider a Wigner function representing a state with a coherence length $l_c$ interacting with a harmonic potential with period $L$. The sections of the Wigner function in the momentum dimension have the shape of Gaussian peaks of width $\hbar/l_c$. As a consequence, the derivatives ${\partial^n W_{\eta\eta}}/{\partial p^n}$ scale with $(l_c)^{n}$, while the spatial derivatives of the harmonic potential scale with $L^{-n}$. Together, this leads to scaling of the terms in \eqref{eq-pot-evolution-commuting} as
\begin{equation}
\frac{1}{n!}\left(\frac{\hbar}{2i}\right)^n\frac{\partial^n W_{\eta\xi}}{\partial p^n}\frac{\partial^n U_{\eta\eta}}{\partial x^n} = O\left(\frac{{l_c}^n}{n!\,L^n}\right).
\end{equation}
Therefore, higher orders can be safely neglected if the length scale of the potential variations is much greater than the coherence length of the initial quantum state. Truncating the series \eqref{eq-pot-evolution-commuting} to first order, one obtains
\begin{equation} 
\left[\dot W_{\eta\xi}(x,p)\right]_U =  W_{\eta\xi}\frac{U_{\eta\eta}-U_{\xi\xi}}{i\hbar}-\frac{\partial W_{\eta\xi}}{\partial p}\frac{F_{\eta\eta}+F_{\xi\xi}}{2},
\end{equation}
where $F_{\eta\eta}(x)= - \partial U_{\eta\eta}/\partial x$ is the force acting on the quantum state $\eta$. This is Eqn.6 in the paper.
Together with the evolution due to kinetic energy, this can be integrated formally for small time steps $\delta t$ to 
\begin{align}
W_{\eta\xi}(x,p;t+\delta t)  = &\\ \nonumber
 e^{-i\frac{U_{\eta\eta}-U_{\xi\xi}}{\hbar}\delta t} W_{\eta\xi}(x-\frac{p}{m}\delta t, p-\frac{F_{\eta\eta}+F_{\xi\xi}}{2}\delta t; t). &
\end{align}

If the potential energy is not diagonal in the chosen basis, the corresponding expression is
\begin{equation} 
\left[\dot W_{\eta\xi}(x,p)\right]_U = \frac{1}{i\hbar}\left(U_{\eta\zeta}W_{\zeta\xi}-W_{\eta\zeta}U_{\zeta\xi} \right)-\frac{1}{2}\left( F_{\eta\xi}\frac{\partial W_{\zeta\xi}}{\partial p} + \frac{\partial W_{\eta\zeta}}{\partial p} F_{\zeta\xi}\right).
\end{equation}
Even though we have used a scalar notation for the position and momentum, the expressions do not change significantly if several spatial dimensions are taken into consideration. In that case, the derivatives become gradients in position and momentum space, respectively. The full equation of motion truncated to first order then reads
\begin{equation} 
\dot W_{\eta\xi} = -\frac{1}{m}\tensor{p}\cdot\nabla_{\!\tensor{x}}W_{\eta\xi}+ \frac{1}{i\hbar}\left(U_{\eta\zeta}W_{\zeta\xi}-W_{\eta\zeta}U_{\zeta\xi} \right)-\frac{1}{2}\left( \tensor{F}_{\!\eta\xi}\cdot \nabla_{\!\tensor{p}} W_{\zeta\xi}+\nabla_{\!\tensor{p}} W_{\eta\zeta}\cdot \tensor{F}_{\!\zeta\xi}\right).
\end{equation}

\begin{equation}\label{eq-sg-Waabb-symbolic}
\begin{array}{l}
W_{\alpha\alpha} = \frac{1}{2}W_0(x-\dfrac{p}{m} t-\dfrac{\gamma\hbar G_{xy}}{4 m} t^2,p+\dfrac{\gamma\hbar G_{xy}}{2}t) \\[4mm]
W_{\beta\beta} = \frac{1}{2}W_0(x-\dfrac{p}{m} t+\dfrac{\gamma\hbar G_{xy}}{4 m} t^2,p-\dfrac{\gamma\hbar G_{xy}}{2}t)
\end{array}
\end{equation}

A similar derivation to that above, starting from (\ref{eq-WdotU}), can be used to express the potential-energy part of the time derivative in terms of the spatial frequency components of the potential. The resulting expression, which may well be more useful than the above when dealing with periodic potentials, such as those generated by standing light waves or similar, is
\begin{equation}
 \left[ \dot{W}_{\eta \epsilon}(x,p)\right]_{U} = \frac{1}{i\hbar} \int \tilde{U}_{\epsilon \rho}(\omega) e^{i\omega x} W_{\eta \rho}^{*}(x,p+\hbar \omega/2) - \tilde{U}_{\eta \rho}(\omega) e^{i\omega x} W_{\rho \epsilon}^{*}(x,p-\hbar \omega/2) ~d\omega, 
\end{equation} 
where we have defined 
\begin{equation}
 U_{\eta \epsilon}(x) = \int e^{i \omega x} \tilde{U}_{\eta \epsilon}(\omega) ~d\omega.
\end{equation}
Once again, this simplifies significantly in the case of a diagonal spin Hamiltonian, and in the case of potentials with only a finite number of spatial frequency components the integral can be replaced with a summation.

\section{Gaussian Beam Profile}
\begin{equation}
W_0(x,p)=\frac{2\sqrt{\ln 2} l_c}{\pi h \,\Delta x}
\exp\left({-\frac{x^2}{\Delta x^2}-\frac{4 l_c^2 p^2 \ln 2}{h^2}}\right).
\end{equation}

\section{Potential Energy Matrices for the Stern-Gerlach and Rabi Experiments}
Fig.3 B and C in the paper show the spatial and momentum traces of the beam passing through the Rabi apparatus in the absence of a radio frequency field. The profile has been obtained by direct numerical integration of the equations of motion \eqref{eq-sg-diagonal-potential}. 
Using \eqref{eq-pot-evolution-commuting}, the equations of motion become
\begin{equation}
\begin{array}{l}
\dot{W}_{\alpha\alpha}=-\dfrac{p}{m}\dfrac{\partial W_{\alpha\alpha}}{\partial x} 
+ \dfrac{\gamma\hbar}{2} G_{xy}  \dfrac{\partial W_{\alpha\alpha}}{\partial p} \\[3mm]
\dot{W}_{\beta\beta}=-\dfrac{p}{m}\dfrac{\partial W_{\beta\beta}}{\partial x} 
- \dfrac{\gamma\hbar}{2} G_{xy} \dfrac{\partial W_{\beta\beta}}{\partial p} \\[3mm]
\dot{W}_{\alpha\beta}=-\dfrac{p}{m}\dfrac{\partial W_{\alpha\beta}}{\partial x}-i \gamma (B_{y0}+x\,G_{xy})W_{\alpha\beta}  \\[3mm]
\dot{W}_{\beta\alpha}=-\dfrac{p}{m}\dfrac{\partial W_{\beta\alpha}}{\partial x}+i \gamma (B_{y0}+x\,G_{xy})   W_{\beta\alpha}. \\[3mm]
\end{array}\label{eq-sg-diagonal-potential}
\end{equation}

In a homogeneous magnetic field gradient, the two spin states experience opposite curvatures. However, due to the symmetry of the apparatus, both spin states pass through the collimation slits centred at $x=0$ at the ends of the gradients. This situation changes if the spin state is perturbed by rf irradiation in the homogeneous section, as shown in Fig.3 B in the paper. In the interest of clarity, the problem has been slightly simplified, by choosing $B_x^0=0$ and $\omega_{rf}=0$. In this variant, the spins precess around a static field in the $z$ direction. Since the field in this section of the experiment is homogeneous, the time evolution is governed completely by the zero order potential energy terms. The potential energy matrix becomes
\begin{equation}
\begin{array}{cc}
U_{\alpha\alpha}(x) = 0 & 
U_{\alpha\beta}(x)  = -\frac{\gamma\hbar}{2}B_1  \\[2mm]
U_{\beta\alpha}(x) = -\frac{\gamma\hbar}{2}B_1   & 
U_{\beta\beta}(x)  = 0 , 
\end{array}
\end{equation}
and the equations of motion are given by
\begin{equation}
\begin{array}{l}
\dot{W}_{\alpha\alpha}=-\dfrac{p}{m}\dfrac{\partial W_{\alpha\alpha}}{\partial x}-\dfrac{i \gamma B_1}{2}( W_{\beta\alpha}-W_{\alpha\beta})  \\[3mm]
\dot{W}_{\beta\beta}=-\dfrac{p}{m}\dfrac{\partial W_{\beta\beta}}{\partial x}-\dfrac{i \gamma B_1}{2}( W_{\alpha\beta}-W_{\beta\alpha})  \\[3mm]
\dot{W}_{\alpha\beta}=-\dfrac{p}{m}\dfrac{\partial W_{\alpha\beta}}{\partial x} -\dfrac{i \gamma B_1}{2}( W_{\beta\beta}-W_{\alpha\alpha})\\[3mm]
\dot{W}_{\beta\alpha}=-\dfrac{p}{m}\dfrac{\partial W_{\beta\alpha}}{\partial x} -  \dfrac{i \gamma B_1}{2}( W_{\alpha\alpha}-W_{\beta\beta}).\\[3mm]
\end{array}
\end{equation}

The resulting Larmor precession is clearly visible in the projections of the components $W_{\alpha\alpha}$ and $W_{\beta\beta}$ (Fig.3 C and D in the paper).
\end{widetext}

\bibliography{spin-scattering,wigner-spin}
\end{document}